\documentclass[12pt,preprint]{aastex}

\shorttitle{Sparse Faraday Rotation Measure Synthesis}
\shortauthors{Andrecut et al.}

\begin{document}


\title{Sparse Faraday Rotation Measure Synthesis}


\author{M. Andrecut, J. M. Stil and A. R. Taylor}
\affil{Institute for Space Imaging Science,\\ Department of Physics and Astronomy,\\ University of Calgary, Calgary, Alberta, T2N 1N4, Canada}



\begin{abstract}
Faraday rotation measure synthesis is a method for analyzing multichannel
polarized radio emissions, and it has emerged as an important tool
in the study of galactic and extra-galactic magnetic fields. The method
requires the recovery of the Faraday dispersion function from measurements
restricted to limited wavelength ranges, which is an
ill-conditioned deconvolution problem. Here, we discuss a recovery
method, which assumes a sparse approximation of the Faraday dispersion
function in an over-complete dictionary of functions. We discuss the
general case, when both thin and thick components are included in the
model, and we present the implementation of a greedy deconvolution
algorithm. We illustrate the method with several numerical
simulations that emphasize the effect of the covered range and sampling resolution
in the Faraday depth space, and the effect of noise on the observed
data.
\end{abstract}


\keywords{Methods: data analysis - Techniques: polarimetric - magnetic fields}



\section{Introduction}

Faraday rotation is a physical phenomenon where the position angle
of linearly polarized radiation propagating through a magneto-ionic
medium is rotated as a function of frequency. The work on astrophisical Faraday
rotation has been initiated in \citep{Burn1966}, and since then several
important contributions have been added to this topic \citep{Gardner1966,Sokoloff1998,Sokoloff1999,Kronberg1994,Vallee1980,Widrow2002}.
Recently, Faraday rotation measure (RM) synthesis has been re-introduced
as an important method for analyzing multichannel polarized radio
data, where multiple emitting regions are present along the single
line of sight of the observations \citep{Brentjens2005,Heald2009}. In practice,
the method requires the recovery of the Faraday dispersion function
from measurements restricted to limited wavelength ranges, which is
an ill-conditioned deconvolution problem, raising important computational
difficulties. Since then, three different approaches have been proposed
to solve this problem. A first approach uses an adaptation of the
Hogbom CLEAN algorithm \citep{Hogbom1974} to the RM deconvolution \citep{Heald2009}.
The second approach is wavelet-based, and assumes field symmetries
in order to project the observed data onto $\lambda^{2}<0$ \citep{Frick2010}.
The third approach \citep{Wiaux2009,Li2011} is based on the compressed
sensing paradigm \citep{Donoho2006,Candes2006}. All these methods are more or
less successful in the case of mixed problems, i.e. when both thin
and thick components are included in the model. For example, in a recent
paper it has been shown that RM Synthesis may yield an erroneous Faraday
structure in the presence of multiple, interfering RM components,
even when cleaning of the Faraday spectrum is performed \citep{Farnsworth2011}.
Also, to our knowledge these methods have not been evaluated in the
presence of noise added to the observed data, a situation that makes
the deconvolution problem even more difficult. Thus, the development
of robust deconvolution methods for the recovery of the Faraday dispersion
function in a given spectral range becomes crucial for the RM synthesis
applications.

Inspired by the above mentioned contributions, in this paper we discuss
the case of sparse approximation of the complex Faraday dispersion
function, i.e. we assume that $F(\phi)$ can be approximated by a
small number of discrete components, which can be both thin or thick. 
Also, we present the implementation of a greedy deconvolution algorithm,
and we illustrate the described method with several numerical simulations
which emphasize the effect of the covered range and sampling resolution in
the Faraday depth space, and the effect of noise on the observed data.
The numerical results show that the described method performs quite
well for simple component mixtures, at typical sampling resolution values and
coverage range in the Faraday depth space, and it is quite robust
in the presence of noise. We show that the described technique is
well suited for exploratory data analysis, where prior information
about the component distributions is not available, and it can be used
as a complement to the previously proposed methods.

Although a sparse solution is an idealized model of a complex
astrophysical system, the potential complexity of the solutions is
adequate for a wide range of astrophysical situations. The sparseness
requirement steers the solution to include the smallest number of
components required to fit an observed Faraday depth spectrum.
Double-lobed radio galaxies that are not resolved by the telescope may
experience different Faraday rotation in each lobe because the
differences in the foreground on scales smaller than the beam. The
lobes themselves may be extended and experience differential Faraday
rotation as well. A sparse solution may consist of two discrete
Faraday components representing each lobe. If the data are good enough
to detect differential Faraday rotation across the source, the
solution may include one or more components with a finite extent in
Faraday depth. Complex source structure may be built up out of a
dictionary of basic thin and thick Faraday components, subject to the
requirement that the solution remains sparse.
 
In the diffuse interstellar medium, a case where Faraday rotation of
Galactic synchrotron emission is dominated by a single HII region
along the line of sight is an example of a system that is well
approximated with two components in Faraday depth, e.g. the circular
Faraday screen discussed by \citet{haverkorn2003} and
\citet{debruyn2009}. As in the case of double lobed radio sources, the
sparse solution is not limited by two delta functions in Faraday
depth, as it can increase in complexity if warranted by the data.

The assumption of sparseness may fail in case there is
a power on a large range of Faraday depths, defined by the minimum and
maximum Faraday depth detectable in a survey. This may occur in some
supernova remnants with complex structure and strong magnetic fields.

\section{Rotation measure synthesis}

In this section we give a brief description of the Faraday RM synthesis
problem, following the formulation introduced in \citet{Brentjens2005}.

The Faraday rotation is characterized by the Faraday depth (in $\mathrm{rad}\,\mathrm{m}^{-2}$),
which is defined as:\begin{equation}
\phi(r)=0.81\int_{source}^{observer}n_{e}B\cdot dr,\end{equation}
 where $n_{e}$ is the electron density (in $cm^{-3}$) , $B$ is
the magnetic field (in $\mu G$), and $dr$ is the infinitesimal path
length (in parsecs). We also define the complex polarization as:\begin{equation}
P(\lambda^{2})=Q(\lambda^{2})+iU(\lambda^{2})=pIe^{2i\chi(\lambda^{2})},\end{equation}
 where $p$ is the fractional polarization, $I$, $Q$, $U$ are the Stokes parameters, and $\chi(\lambda^{2})$ is the polarization
angle observed at wavelength $\lambda$:\begin{equation}
\chi(\lambda^{2})=\frac{1}{2}\arctan\frac{U(\lambda^{2})}{Q(\lambda^{2})}.\end{equation}
The Faraday RM is defined as the derivative of the polarization angle $\chi(\lambda^{2})$, with respect to $\lambda^{2}$:
\begin{equation}
RM(\lambda^{2})=\frac{d\chi(\lambda^{2})}{d\lambda^{2}}.\end{equation}

We now identify RM with the Faraday depth $\phi$, and we assume that
the observed polarization $P(\lambda^{2})$ originates from the emission
at all possible values of $\phi$, such that:
\begin{equation}
P(\lambda^{2})=\int_{-\infty}^{+\infty}F(\phi)e^{2i\phi\lambda^{2}}d\phi,\end{equation}
 where $F(\phi)$ is the complex Faraday dispersion function (the
intrinsic polarized flux, as a function of the Faraday depth). Thus,
in principle $F(\phi)$ is the inverse Fourier transform of the observed
quantity $P(\lambda^{2})$:\begin{equation}
F(\phi)=\int_{-\infty}^{+\infty}P(\lambda^{2})e^{-2i\phi\lambda^{2}}d\lambda^{2}.\end{equation}
 However, this operation is ill-defined since we cannot observe $P(\lambda^{2})$
for $\lambda^{2}<0$, and also in practice the observations are limited
to an interval $[\lambda_{min}^{2},\lambda_{max}^{2}]$.

In order to deal with the above limitations, the observed polarization
is defined as:\begin{equation}
\tilde{P}(\lambda^{2})=W(\lambda^{2})P(\lambda^{2}),\end{equation}
 where $W$ is the observation window function, with $W(\lambda^{2})>0$
for $\lambda^{2}\in[\lambda_{min}^{2},\lambda_{max}^{2}]$, and $W(\lambda^{2})=0$
otherwise. Therefore, we obtain the reconstructed dispersion function:\begin{equation}
\tilde{F}(\phi)=A\int_{-\infty}^{+\infty}\tilde{P}(\lambda^{2})e^{-2i\phi\lambda^{2}}d\lambda^{2},\end{equation}
 where \begin{equation}
A=\left[\int_{\lambda_{min}^{2}}^{\lambda_{max}^{2}}W(\lambda^{2})d\lambda^{2}\right]^{-1},\end{equation}
 is the normalization constant for the observation window. The reconstructed
dispersion function can also be written as:\begin{equation}
\tilde{F}(\phi)=R(\phi)\circ F(\phi),\end{equation}
 where $\circ$ is the convolution operator, and\begin{equation}
R(\phi)=A\int_{-\infty}^{+\infty}W(\lambda^{2})e^{-2i\phi\lambda^{2}}d\lambda^{2},\end{equation}
 is the RM spread function (RMSF).

Using the shift theorem, we can also write:\begin{equation}
\tilde{F}(\phi)=R(\phi)\circ F(\phi)=A\int_{-\infty}^{+\infty}\tilde{P}(\lambda^{2})e^{-2i\phi(\lambda^{2}-\bar{\lambda}^{2})}d\lambda^{2},\end{equation}
 and\begin{equation}
R(\phi)=A\int_{-\infty}^{+\infty}W(\lambda^{2})e^{-2i\phi(\lambda^{2}-\bar{\lambda}^{2})}d\lambda^{2}.\end{equation}
 where $\bar{\lambda}^{2}$ is the mean of the sampled values in $[\lambda_{min}^{2},\lambda_{max}^{2}]$.

The goal of the analysis is to find $F(\phi)$ from the observed values
$\tilde{P}(\lambda_{n}^{2})=\tilde{P}_{n}$ (i.e. $\tilde{Q}_{n}$
and $\tilde{U}_{n}$) over $N$ discrete channels $\lambda_{n}^{2}\in[\lambda_{min}^{2},\lambda_{max}^{2}]$,
$n=0,1,...,N-1$, with the given weights $W(\lambda_{n}^2)=W_{n}$.
Since the measured values are discrete (each value constitutes an
integral over the channel centered at $\lambda_{n}^{2}$), we should
consider the discrete versions of the above equations, i.e.:
\begin{equation}
\tilde{F}(\phi)\simeq A\sum_{n=0}^{N-1}\tilde{P}_{n}e^{-2i\phi(\lambda_{n}^{2}-\bar{\lambda}^{2})},\end{equation}
 and respectively \begin{equation}
R(\phi)\simeq A\sum_{n=0}^{N-1}W_{n}e^{-2i\phi(\lambda_{n}^{2}-\bar{\lambda}^{2})}.\end{equation}
 The reconstructed function $\tilde{F}(\phi)$ depends on the window
$W(\lambda^{2})$, which acts as a filter, and improves substantially
by increasing its coverage in the $\lambda^{2}$ space. Obviously,
$\tilde{F}(\phi)$ is a {}``dirty'' reconstruction of $F(\phi)$,
i.e. the convolution of $F(\phi)$ with $R(\phi)$, and and a deconvolution
step is necessary to recover $F(\phi)$.

\section{Sparse approximation}

\subsection{Discrete representation}

In general, the number of data points is limited by the number of
independent measurement channels, and therefore there are many different
potential Faraday dispersion functions consistent with the measurements
\citep{Burn1966,Brentjens2005,Heald2009,Frick2010,Li2011,Farnsworth2011}. The usual approach
to resolving such ambiguities, is to impose some extra constraints
on the Faraday dispersion function. 
Our approach is based on the recently introduced framework of compressive sensing \citep{Donoho2006,Candes2006}.
Compressive sensing relies on the observation that many types of signals
can be well-approximated by a sparse expansion in terms of a suitable
basis, or dictionary of functions. The main idea of compressive sensing is that if
the signal is sparse, then a small number of measurements contain
sufficient information for its approximate or exact recovery. In our
case, the problem is to reconstruct a sparse $F(\phi)$ from a relatively
small number of $\tilde{P}(\lambda^{2})$ measurements.
Therefore, we assume that the model of $F(\phi)$
is sparse in an over-complete dictionary of functions. By over-complete
we understand that the number of functions in the dictionary is larger
than the number of independent observation channels. Thus,
the dictionary functions may be redundant (linearly dependent), and
therefore non-orthogonal. In order to give a proper formulation of
this approach we need to introduce a discrete representation of the
$\phi$ space. 

It is known \citep{Brentjens2005} that, for a discrete sampled Faraday dispersion function, 
the full width at half maximum of the main peak of the RMSF is given by:
\begin{equation}
\delta\phi=\frac{2\sqrt{3}}{\Delta\lambda^{2}},
\end{equation}
where $\Delta\lambda^{2}$ is the width of the observation interval. 
Also, using a uniform grid in $\lambda^{2}$ space one can estimate the maximum observable Faraday depth by:
\begin{equation}
\phi_{max}=\frac{\sqrt{3}}{\delta\lambda^{2}},
\end{equation}
where $\delta\lambda^{2}=\Delta\lambda^{2}/N$ is the width of an observing channel \citep{Brentjens2005}.
This estimation of $\phi_{max}$ is only an approximation, since in reality only the frequency $\nu$
is sampled linearly. Therefore, in our discrete representation we consider 
a nonlinear grid in the $\lambda^{2}$ space: $\lambda_{n}^{2}=c^{2}/\nu_{n}^{2}$,
where $\nu_{n}=(\nu_{max}-\nu_{min})/N$ is the centered frequency
of the channel $n=0,1,...,N-1$, and $c$ is the speed of light. 
Also, we consider a linear grid in the $\phi$ space, where the computational window $\phi_{win}$, 
the sampling resolution $\phi_{R}$, and the number of points $M$ are set to: 
\begin{equation}
\phi_{win}\leq\phi_{max},\quad\phi_{R}\leq\delta\phi,\quad M=\left\lfloor \frac{\phi_{win}}{\phi_{R}}\right\rfloor ,
\end{equation}
where $\left\lfloor x\right\rfloor $ is the integer part of $x$. 

The model of $F(\phi)$ is therefore characterized by a uniform grid, $\phi_{m}=-\phi_{win}+m\phi_{R}$, $m=0,1,...,M-1$, 
and a vector $z=[z_{0},z_{1},...,z_{M-1}]\in\mathbb{\mathbb{C}}^{M}$, which 
is assumed sparse, i.e. it has a small number of non-zero components, 
corresponding to the complex amplitudes of the sources located on the $\phi_{m}$
grid. For example, a thin source with the amplitude $z_{m}$, located
at $\phi_{m}$ , will be approximated by the product of $z_{m}$ with
a Dirac function $\delta(\phi-\phi_{m})$, while a thick source will
be characterized by a contiguous set of non-zero amplitudes in the
vector $z$, which requires a different set of adaptive functions, capable
of capturing their position and extensive support in the $\phi$ space. 
The goal of the analysis is to find the vector
$z$, which is a discrete approximation of the Faraday dispersion
function $F(\phi)$, from the measurements $\tilde{Q}_{n}$ and $\tilde{U}_{n}$,
$n=0,1,...,N-1$.

\subsection{Dirac approximation}

Since, in general we can have $M\geq N$, the Dirac functions $\delta(\phi-\phi_{m})$,
$m=0,1,...,M-1$, form an over-complete dictionary in the $\phi$
space. The decomposition of $F(\phi)$ with respect
to the Dirac over-complete dictionary is:\begin{equation}
F(\phi)=\sum_{m=0}^{M-1}z_{m}\delta(\phi-\phi_{m}).\end{equation}
From the equations (5) and (7) we obtain: \begin{equation}
\tilde{P}(\lambda^{2})=W(\lambda^{2})\int_{-\infty}^{+\infty}\sum_{m=0}^{M-1}z_{m}\delta(\phi-\phi_{m})e^{2i\phi\lambda^{2}}d\phi=W(\lambda^{2})\sum_{m=0}^{M-1}z_{m}e^{2i\phi_{m}\lambda^{2}}.\end{equation}
We observe that the transformation of $F(\phi)$ into $\tilde{P}(\lambda^{2})$
can be written in a matrix form as following:\begin{equation}
W\Psi z=\tilde{p},\end{equation}
 where \begin{equation}
\tilde{p}=[\tilde{P}_{0},\tilde{P}_{1},...,\tilde{P}_{N-1}]^{T}\in\mathbb{C}^{N},\end{equation}
 is the $N$-dimensional complex vector of observations, and $\Psi\in\mathbb{C}^{N\times M}$
is the $N\times M$ matrix with the Fourier terms:\begin{equation}
\Psi_{n,m}=e^{2i\phi_{m}\lambda_{n}^{2}},\end{equation}
 and $W$ is the $N\times N$ diagonal matrix, with the diagonal elements
equal with the channel weights: $W_{n,n}\equiv W_{n}$. 

If we are searching for the sparsest solution possible, then the $\ell_{0}$
norm of $z$:
\begin{equation}
\left\Vert z\right\Vert _{0}=\sum_{m=0}^{M-1}h(z_{m}),\end{equation}
\begin{equation}
h(z_{m})=\left\{ \begin{array}{ccc}
1 & if & \left|z_{m}\right|>0\\
0 & if & z_{m}=0\end{array}\right.,\end{equation}
 should be minimized. This sparseness assumption leads to the following
optimization problem:\begin{equation}
\min_{z}\left\Vert z\right\Vert _{0}\quad subject\: to\quad W\Psi z=\tilde{p}.\end{equation}
However, finding the minimum $\ell_{0}$ norm is an NP-complete problem,
which requires a combinatorial search of the parameter space and therefore
is practically unfeasible \citep{Donoho2006,Candes2006}. A better approach is
to replace the $\ell_{0}$ norm with the $\ell_{1}$ norm:\begin{equation}
\left\Vert z\right\Vert _{1}=\sum_{m=0}^{M-1}\left|z_{k}\right|,\end{equation}
which transforms the combinatorial problem into a convex problem,
that can be solved in polynomial time \citep{Boyd2004}, and it has been
shown to give solutions close to the $\ell_{0}$ norm solutions \citep{Chen2001}.
Thus, the problem can be reformulated as finding the vector $z$ such
that:\begin{equation}
\min_{z}\left\Vert z\right\Vert _{1}\quad subject\: to\quad W\Psi z=\tilde{p}.\end{equation}
One can see that we do not make any assumption on the number of non-zero
components, we just assume that their number is smaller than $M$.

So far we have not considered the influence of noise on the observed
data. We assume a complex noise vector $\eta\in\mathbb{C}^{N}$, with
the components $\eta_{n}\in\mathbb{C}$ having the real and respective
imaginary parts sampled from a normal distribution with zero mean
and standard deviation $\sigma$: $\mathrm{Re}\{\eta_{n}\},\mathrm{Im}\{\eta_{n}\}\in N(0,\sigma)$.
Thus, the transformation of $F(\phi)$ into $\tilde{P}(\lambda^{2})$
can be rewritten as:\begin{equation}
W\Psi z+\eta=\tilde{p},\end{equation}
 and the minimization problem can be reformulated as:\begin{equation}
\min_{z}\left\Vert z\right\Vert _{1}\quad subject\: to\quad\left\Vert W\Psi z-\tilde{p}\right\Vert _{2}^{2}\leq(\beta\sigma)^{2}.\end{equation}
The use of the $\ell_{1}$ norm induces sparsity in $z$, while the
constraint ensures $W\Psi z\approx\tilde{p}$. Since $\tilde{p}$
is observed in the presence of noise, it is reasonable to not enforce
$W\Psi z=\tilde{p}$ exactly, and to stop the minimization process
when the norm of the residual becomes comparable with the standard
deviation of the noise ($\beta\sim\sqrt{N}$).

\subsection{Generalization}

Dirac functions can be used to approximate thin sources only. In orderDonoho2006
to approximate thick sources we extend the dictionary by incorporating
a set of functions, characterized by adaptive translation and scaling
properties, such that they are capable to capture the position and
the extent of thick sources in the $\phi$ space. Thus, we assume
that $F(\phi)$ has a sparse approximation in an over-complete dictionary
$\Phi$ of functions $\varphi_{j}(\phi)\in\Phi$, called atoms \citep{Mallat1993}:
\begin{equation}
F(\phi)=\sum_{j=0}^{J}\xi_{j}\varphi_{j}(\phi).\end{equation}
 Here, $J\geq M$ is the number of atoms in the dictionary, and only
a small number of the complex coefficients $\xi_{j}$ are assumed
to be non-zero. Thus, by introducing the $M\times J$ complex matrix
$\Phi\in\mathbb{C}^{M\times J}$, with the elements $\Phi_{m,j}=\varphi_{j}(\phi_{m})$,
and the sparse complex vector $\xi=[\xi_{0},\xi_{1},...,\xi_{J-1}]^{T}\in\mathbb{C}^{J}$,
and taking into account that: \begin{equation}
\Phi\xi=z,\end{equation}
 we obtain the following minimization problem:\begin{equation}
\min_{\xi}\left\Vert \xi\right\Vert _{1}\quad subject\: to\quad\left\Vert \Gamma\xi-\tilde{p}\right\Vert _{2}^{2}\leq(\beta\sigma)^{2},\end{equation}
where\begin{equation}
\Gamma=W\Psi\Phi,\end{equation}
is a $N\times J$ complex matrix. In this more general case, the goal
is to find the sparse vector $\xi\in\mathbb{C}^{J}$ in the dictionary
space. Obviously, when the dictionary is reduced to the Dirac basis
we have $J=M$, $\xi\equiv z$ and $\Phi\equiv I$, where $I$ is
the $M\times M$ identity matrix, and therefore $\Gamma$ reduces
to the weighted Fourier matrix, $\Gamma=W\Psi$.

\subsection{Over-complete dictionaries}

We should note that an over-complete dictionary $\Phi$ that leads
to sparse representations can be chosen as a pre-specified set of
analysis functions (wavelets, Gaussian packets, Gabor functions etc.),
or designed by modeling its content to a given set of signal examples
\citep{Candes2006,Mallat1993}. The success of such dictionaries in applications
depends on how suitable they are to sparsely describe the signals
in question. A general family of analysis functions can be obtained
by scaling and translating a single normalized window function $\varphi(\phi)$,
with $\left\Vert \varphi\right\Vert _{2}=1$. Therefore, for any scale
$a>0$ and translation $b\in\mathbb{R}$ we define the atom $\varphi_{j}$
of the dictionary $\Phi$ as following:\begin{equation}
\varphi_{j}(\phi)\equiv\varphi_{j(a,b)}(\phi)\equiv\frac{1}{\sqrt{a}}\varphi\left(\frac{\phi-b}{a}\right).\end{equation}
Therefore, the index $j$ of the atom function depends on both $a$
and $b$ parameters: $j=j(a,b)$. Thus, in order to represent $F(\phi)$
in the dictionary $\Phi$, we need to select an appropriate countable
subset of atoms $\varphi_{j}$, $j=0,1,...,J-1$, such that $F(\phi)$
can be written as a linear expansion. Depending on the choice of the
atoms $\varphi_{j}$, the expansion coefficients will give explicit
information about the behavior of $F(\phi)$. For example, we should
note here that different wavelet transforms correspond to different
families of atoms. In our definition, we do not limit the dictionary
to a single wavelet basis, on contrary we consider an over-complete
set, which also may contain different concatenated families (sub-dictionaries)
of such analysis functions. In order to illustrate numerically this
approach, let us consider the boxcar dictionary, defined as: \begin{equation}
\varphi_{j(a,b)}(\phi)=\left\{ \begin{array}{ccc}
1/\sqrt{a} & if & b\leq\phi<b+a\\
0 &  & otherwise\end{array}\right..\end{equation}
An important characteristic of the boxcar dictionary is that it can
capture sources with arbitrary thickness. Another advantage is its
easy discretization. In our case, the discretization grid has $M$
points $\phi_{m}$ with the sampling resolution $\phi_{R}$. Thus, assuming
that the maximum width of a boxcar atom is $a_{max}=S\phi_{R}$, where
$S\leq\left\lfloor M/2\right\rfloor $, then for each scale $a=s\phi_{R}$,
$s=1,2,...,S$, and translation $b=l\phi_{R}$, $l=0,1,...,M-s$
we can define a boxcar function with the index $j=j(s,l)$, such that:\begin{equation}
\varphi_{j}(\phi_{m})\equiv\varphi_{j(s,l)}(\phi_{m})=\left\{ \begin{array}{ccc}
1/\sqrt{s\phi_{R}} & if & l\leq m<l+s\\
0 &  & otherwise\end{array}\right..\end{equation}
Therefore, one can define maximum $J=SM-S(S+1)/2$ boxcar functions
on such a grid, and we can easily build a discrete dictionary matrix
$\Phi$ of size $M\times J$. In this paper we limit our discussion
to the boxcar dictionary defined above, since it is simple enough
to illustrate the approach, and to provide meaningful results. Also,
this dictionary includes by construction the Dirac set of functions,
which in this case are the first $M$ functions with $s=1$.
A similar approach can be used to build sub-dictionaries corresponding
to other families of analysis functions.

\subsection{Multi-scale analysis}

The sparse decomposition can also be used to perform a multi-scale
analysis, by considering all the dictionaries $\Phi_{S}$, where $S=1,2,...,S_{max}\leq\left\lfloor M/2\right\rfloor $.
Also, let us assume that $z_{S}=\Phi_{S}\xi$ is the solution obtained
for the scale $S$, i.e. the recovered discrete representation of
$F(\phi)$ with the dictionary $\Phi_{S}$. We consider a $S_{max}\times M$
matrix $\Xi$, where each line with the index $S$ corresponds to
the solution obtained for the scale $S$, i.e. $\Xi_{S}\equiv z_{S}=\Phi_{S}\xi$.
Obviously, the solution $z_{S}$ will depend on the maximum scale
$S$ used in each dictionary $\Phi_{S}$, and by visualizing the matrix
$\Xi$, we obtain a representation of the behavior of the solution
at different scales.

\section{Matching pursuit}

The sparse optimization problem, defined in the previous section,
is known as Basis Pursuit Denoising (BPD) \citep{Donoho2006}, and if written
in a Lagrangian form: \begin{equation}
\min_{\xi}\left[\frac{1}{2}\left\Vert \Gamma\xi-\tilde{p}\right\Vert _{2}^{2}+\alpha\left\Vert \mathbf{\xi}\right\Vert _{1}\right],\end{equation}
 it can be thought of as a least squares problem with an $\ell_{1}$
regularizer, where $\alpha>0$ is a parameter that controls the trade-off
between sparsity and reconstruction fidelity. Thus, BPD solves a regularization
problem with a trade-off between having a small residual and making
the solution simple in the $\ell_{1}$ sense. The solutions of BPD
are often the best computationally tractable approximation of the
under-determined system of equations \citep{Donoho2005}. In our case, since
the direct space and the inverse Fourier space are perfectly incoherent,
the problem can be solved using linear programming techniques whose
computational complexities are polynomial. However, for the sparse
RM approximation problem, the BPD approach requires the solution of
a very large convex, non-quadratic optimization problem, and therefore
suffers from high computational complexity. Due to the complexity
of the linear programming approach, several other $\ell_{1}$ optimization
methods have been proposed to solve the BPD problem \citep{Donoho2006,Candes2006}.
Here, we consider a method based on sub-optimal greedy algorithms,
which requires far less computation. Our goal is not only to obtain
a good sparse expansion, but also to provide a fast computational
method, therefore here we focus our attention on the greedy Matching
Pursuit (MP) algorithm \citep{Mallat1993}, which is the fastest known
algorithm for the BPD problem \citep{Chen2001}. MP has many applications
in signal and image coding, shape representation and recognition,
data compression etc. One of its main features is that it can be applied
to arbitrary dictionaries.

Starting from an initial approximation $\xi(0)=0$ and residual $r(0)=\tilde{p}$,
the algorithm uses an iterative greedy strategy to pick the column
vectors $\Gamma^{(j)}$ which best reduce the residual. At every time
step $t$ the current residual $r(t)$ can be decomposed as following:
\begin{equation}
r(t)=\left\langle r(t),\Gamma^{(j)}\right\rangle \left\Vert \Gamma^{(j)}\right\Vert _{2}^{-2}\Gamma^{(j)}+r(t+1),\end{equation}
 where $r(t+1)$ is the future residual, and $\left\langle .,.\right\rangle $
is the standard inner product operator in the complex Hilbert space.
Since $r(t+1)$ and $\Gamma^{(j)}$ are orthogonal, $\left\langle r(t+1),\Gamma^{(j)}\right\rangle =0$,
we have:\begin{equation}
\left\Vert r(t+1)\right\Vert _{2}^{2}=\left\Vert r(t)\right\Vert _{2}^{2}-\left|\left\langle r(t),\Gamma^{(j)}\right\rangle \right|^{2}\left\Vert \Gamma^{(j)}\right\Vert _{2}^{-2}.\end{equation}
 In order to minimize the norm of the future residual, the algorithm
should choose the column vector $\Gamma^{(j)}$ which maximizes the
projection on the current residual:
\begin{equation}
k_{t}=\arg\max_{j}\left\{ \left|\left\langle r(t),\Gamma^{(j)}\right\rangle \right|\left\Vert \Gamma^{(j)}\right\Vert _{2}^{-1}\right\} .\end{equation}
 Therefore, after choosing the best column $\Gamma^{(k_{t})}$ one
can update the solution and the residual as following:\begin{equation}
\xi(t+1)=\xi(t)+c\Gamma^{(k_{t})},\end{equation}
 \begin{equation}
r(t+1)=r(t)-c\Gamma^{(k_{t})},\end{equation}
 where \begin{equation}
c=\left\langle r(t),\Gamma^{(k_{t})}\right\rangle \left\Vert \Gamma^{(k_{t})}\right\Vert _{2}^{-2}.\end{equation}
 Thus, after $t$ iteration steps the resulted solution is a sparse
vector $\xi$ with the non-zero coefficients $\xi_{k_{t}}$. The algorithm
stops when the maximum number of iterations has been reached (which
usually is set to $J$), or when the norm of the residual becomes
comparable with the standard deviation of the noise. The reconstruction
of the target signals is then given by: \begin{equation}
z=\sum_{j=0}^{J-1}\xi_{j}\Phi^{(j)}=\Phi\xi,\end{equation}
 \begin{equation}
\tilde{p}=\sum_{j=0}^{J-1}\xi_{j}\Gamma^{(j)}=\Gamma\xi.\end{equation}
The pseudo-code of the RM-MP algorithm is listed in the Appendix.

\section{Numerical results}

\subsection{Two different experiment layouts}

In order to illustrate the described deconvolution method, we have
considered two different experiment configurations, corresponding to two 
different ranges of observed frequencies. The first one is consistent
with the observations with the Westerbork Synthesis Radio Telescope
(WSRT) in the frequency range 315 MHz to 375 MHz, as described in
\citep{Brentjens2005}. The second one is consistent with the observations
with the Arecibo telescope in the frequency range 1225 MHz to 1525
MHz, for The Galactic ALFA Continuum Survey (GALFACTS), as described
in \citep{Taylor2010}. The separation between the frequency windows is
roughly 1 GHz, and therefore the maximum observable Faraday depth  
and the half maximum of the main peak of the RMSF are quite different.
Here we will show that the RM-MP method provides very good
results in both cases.  

As a testbed for numerical simulations, we have considered a mixed scenario consisting 
of three components with different widths, such that the simulation results provide the response of the algorithm 
to a full range of component widths.  
The first one is a thin component given by: $F(-0.5\phi_{win})=9-8i$.
The second is a thick component given by: $F(\phi)=-7+8i$ if $-0.02\phi_{win}\leq\phi<0.02\phi_{win}$,
and $F(\phi)=0$ otherwise. The third is a thicker component defined
by: $F(\phi)=8-6i$ if $0.46\phi_{win}\leq\phi<0.54\phi_{win}$, and
$F(\phi)=0$ otherwise. Thus, this scenario can be easily scaled for different 
computational windows $[-\phi_{win},\phi_{win}]$, where $\phi_{win}$ is given in $\mathrm{rad}\,\mathrm{m}^{-2}$. 
Also, we have considered that all the observational 
channels are equally weighted: i.e. $W_{n}=1$, $n=0,1,...,N$, and
$A=1/N$.

\subsection{WSRT}

The various parameters
associated with the WSRT experiment layout \citep{Brentjens2005} are listed
bellow:
\medskip{}

Frequency range: $\nu_{min}=315\,\mathrm{MHz}$, $\nu_{max}=375\,\mathrm{MHz}$;

Wave length range: $\lambda_{min}^{2}=0.639\,\mathrm{m}^{2}$, $\lambda_{max}^{2}=0.905\,\mathrm{m}^{2}$,
$\triangle\lambda^{2}=0.266\,\mathrm{m}^{2}$;

Number of channels: $N=126$;

Half maximum of the main peak of the RMSF: $\delta\phi=12.990\,\mathrm{rad}\,\mathrm{m}^{-2}$;

Maximum observable Faraday depth: $\phi_{max}=818.414\,\mathrm{rad}\,\mathrm{m}^{-2}$;

\medskip{}

Let us first consider the ideal noiseless case, when the sampling resolution
in the $\phi$ space is equal with the half maximum of the main peak
of the RMSF, $\phi_{R}=\delta\phi$, and the computational window is $\phi_{win}=\phi_{max}$.
In this particular case, as shown in Figure 1, the RM-MP algorithm
provides an exact solution, since $N=M=126$ and therefore no information
is lost in the measurement. One can also notice that in this noiseless
exact sampling case, the solution is independent of the scale used
in the dictionary, as it can be seen on the multi-scale representation
for $0<S\leq25$. However, the problem becomes ill-defined in the
following situations: the noise is present; the sampling resolution becomes finer than the half maximum of the
main peak of the RMSF, $\phi_{R}<\delta\phi$; and the number of independent
observed channels is smaller than the number of points in the $\phi$
space, $N<M$. In this case the system becomes under-determined, and
therefore some information is lost. In order to exemplify this situation,
we consider a scenario in which all these factors are present. We
add noise with the standard deviation $\sigma=\sqrt{N}=11.22$, to
the $Q$ and $U$ values. We limit the computational window to $\phi_{win}=126\,\mathrm{rad}\,\mathrm{m}^{-2}<\phi_{max}$,
and we increase the number of points on the $\phi$ grid to $M=252$,
which is double of the number of observation channels $N=126$. This
results in a sampling resolution of $\phi_{R}=1\,\mathrm{rad}\,\mathrm{m}^{-2}\ll\delta\phi$.
The obtained results (for $\beta=\sqrt{2N}$, $S=25$) are shown in
Figure 2. One can see that the phase of some components cannot be reliably
recovered anymore, since there is not enough information in the signal
to be detected properly. We should note that the problem
is correctly resolved in the noiseless case (not shown here). Thus,
the effect of noise addition consists in a partial loss of information
about the phase of $F(\phi)$, which is expected, since the number
of solutions compatible with the data increases dramatically with
the added noise. Also, we should point out that the solution improves 
by increasing the signal-to-noise ratio, as shown in Figure 3, where we have 
increased the amplitude of the components by a factor of 1.5, keeping their 
phase unchanged. One can see that in this case, the RM-MP method resolves correctly all
the components. This result suggests that an adequate signal to noise
ratio should be taken into account, in order for the method to be
successful. 

\subsection{Arecibo}

The GALFACTS survey, carried out with the Arecibo telescope, has the following parameters:

Frequency range: $\nu_{min}=1225\,\mathrm{MHz}$, $\nu_{max}=1525\,\mathrm{MHz}$;

Wave length range: $\lambda_{min}^{2}=0.0386\,\mathrm{m}^{2}$, $\lambda_{max}^{2}=0.0598\,\mathrm{m}^{2}$,
$\triangle\lambda^{2}=0.0212\,\mathrm{m}^{2}$;

Half maximum of the main peak of the RMSF: $\delta\phi=163.044\,\mathrm{rad}\,\mathrm{m}^{-2}$;

The maximum observable Faraday depth, $\phi_{max}$, is inverse proportional
with the width of the observation channel $\delta\lambda$, and by
increasing the number of channels, the maximum observable Faraday
depth becomes unreasonable high. Therefore, in order to obtain some meaningful results, we have to limit both the number
of observation channels in the $\lambda^{2}$ space, and the computational window in the $\phi$ space.

First we consider that the compuational window is limited to $\phi_{win}=1800\,\mathrm{rad}\,\mathrm{m}^{-2}$
and the number of observation channels is $N=200$. Also, we consider
the same testbed as for WSRT case, and we add noise with the standard
deviation $\sigma=\sqrt{N}=14.14$. In addition, we increase the number
of points on the $\phi$ grid to $M=300$, and therefore we obtain
a sampling resolution: $\phi_{R}=12\,\mathrm{rad}\,\mathrm{m}^{-2}\ll\delta\phi$.
The obtained results (for $\beta=\sqrt{2N}$, $S=25$) are shown in
Figure 4. One can see that all the components are relatively well resolved,
with a small error in the phase, but with almost exact amplitudes.
In the next experiment we zoom more in the $\phi$ space, and we impose
a computational window of $\phi_{win}=900\,\mathrm{rad}\,\mathrm{m}^{-2}$,
keeping the same number of observation channels and number of points
on the $\phi$ grid, such that $\phi_{R}=6\,\mathrm{rad}\,\mathrm{m}^{-2}$.
The results are again reasonable good, as shown in Figure 5, with
a small variation of the phase due to the uncertainty introduced by the
noise addition. However, if we zoom further in the $\phi$ space the
solution is not as good anymore, as it can be seen in Figure 6. In
this case we have a much finer sampling resolution $\phi_{R}=3\,\mathrm{rad}\,\mathrm{m}^{-2}$,
corresponding to a computational window of $\phi_{win}=600\,\mathrm{rad}\,\mathrm{m}^{-2}$,
number of observation channels $N=400$, and the number of points
on the $\phi$ grid $M=400$. This is a consequence of the fact that
by increasing $N$, we have increased also the standard deviation
of the noise to $\sigma=\sqrt{N}=20$, such that the signal to noise
ratio is smaller than before. Again, an improved solution can be obtained
by increasing the amplitude of the components, such that the signal to noise ratio is higher.

\subsection{Beyond the RMSF resolution}

In the previous numerical experiments we have shown that the RM-MP algorithm is able to resolve correctly the components 
from the input $F(\phi)$ model, if the separation between the components is higher than the half maximum of the main peak of the RMSF. 
In order to estimate the response of the RM-MP algorithm at resolutions beyond the RMSF limit, we consider two Dirac components, 
$F(\phi-\Delta\phi_{in}/2)=9-7i$, and respectively  $F(\phi+\Delta\phi_{in}/2)= -9+7i$, 
separated by $\Delta\phi_{in}<\delta\phi$, where $\delta\phi$ is the half maximum of the main peak of the RMSF. The numerical experiments show that 
the RM-MP algorithm cannot resolve correctly the two components, but returns a boxcar function centered at the exact position value $\phi$,  
with a width equal with the separation between the two components. In order to illustrate this result we consider the WSRT scenario, with 
$\phi_{win}=\phi_{max}$, $N=126$ and $M=4N=1008$, which gives a sampling resolution $\phi_{R}=0.812\,\mathrm{rad}\,\mathrm{m}^{-2}$, in the $\phi$ space. 
In Figure 8 we give the width of the output boxcar function $\Delta\phi_{out}$ as a function of the input separation $\Delta\phi_{in}$. One can see that 
for all performed experiments we have $\Delta\phi_{out} = \Delta\phi_{in}$. Also, in Figure 9 we show a typical example, where the input separation is 
$\Delta\phi_{in}=5\phi_{R}=4.06\,\mathrm{rad}\,\mathrm{m}^{-2}$, or approximatively $30\%$ from $\delta\phi$. Thus, even at resolutions beyond  
the half maximum of the main peak of the RMSF, the RM-MP algorithm provides some useful information, i.e. the position and the separation width of the two components.

\subsection{Discussion}

The above numerical experiments have shown that the sparse RM-MP method works
well for relatively simple sparse problems. We should note that the method can be used 
to recover more complex dispersion functions. For example, let us consider the situation from Figure 7, 
where we have two thin components and two thick components. The first thick component is modeled as a Gaussian, 
while the second is modeled as a boxcar function. Also, we assume the noisy WSRT experiment configuration, with: 
$\sigma=\sqrt{N}$, $\phi_{win}=818.414\,\mathrm{rad}\,\mathrm{m}^{-2}$, $M=220$ and $\phi_{R}=7.440\,\mathrm{rad}\,\mathrm{m}^{-2}<\delta\phi$. 
One can see that all the sources are almost exactly recovered, including the thick Gaussian, even though the dictionary 
does not contain any Gaussian functions. In fact, the shape of the Gaussian is reconstructed from several boxcar functions 
from the dictionary. Thus, the boxcar dictionary can be used to recover more complex functions.
However, the success of the method depends on another aspect which has not yet been discussed. 
More specifically, the performance of the RM-MP method depends on the 
number of observation channels $N$, the number of points $M$ on the $\phi$ grid, and the number $K$
of non-zero components in the discrete representation
of the Faraday depth function $F(\phi)$.
An important question here is that given $N$ and $M$, what is the maximum value of $K$, for a faithful
recovery of $F(\phi)$? In \citep{Donoho2006,Candes2006} it has been shown that
any $K$-sparse signals of length $M$, with $K\ll M$, can be recovered from only
$N\geq cK<M$ random measurements (projections), where $c\sim\log(M/K)$.
The answer to this question is not obvious for the sparse RM synthesis problem, since the reconstruction
process will depend on experiment layout, i.e. the observed frequency band and the half maximum of the main peak of the RMSF. 
This is an important theoretical question which we would like to address in the future development,
in order to improve the performance of the method. 

\section{Conclusions}

The recently introduced Faraday RM synthesis is becoming an important
tool for analyzing multichannel polarized radio data, and derive properties of astrophysical
magnetic fields. The method requires the solution of an ill-conditioned
deconvolution problem, in order to recover the intrinsic Faraday dispersion
function, and therefore the development of robust methods has become
crucial for the RM Synthesis applications. Here, we have assumed that
the complex Faraday dispersion function $F(\phi)$ can be approximated
by a small number of discrete components from an over-complete dictionary,
and we have developed a greedy algorithm to solve the deconvolution
problem. The method uses an over-complete dictionary of functions
which can be efficiently used in a multi-scaling context, and it can
easily include different types of analysis functions. We also have
presented several numerical simulations showing the effect of the
covered range and sampling resolution in the Faraday depth space, and the effect
of noise on the observed data. The numerical results show that the
described method performs well at common resolution values and coverage
range in the Faraday depth space, and it is quite robust in the presence
of noise. Therefore, the described technique is well suited for exploratory
data analysis, and it can be used as a complement to the previously
proposed methods. 

\appendix

\section{Appendix material}

\bigskip
The pseudo-code of the RM-MP algorithm:
\bigskip

$\xi\leftarrow0$; solution vector ($J$-dimensional)

$r\leftarrow\tilde{p}$; initial residual ($N$-dimensional)

$\Gamma\leftarrow W\Psi\Phi$; systems matrix ($N\times J$-dimensional)

$\sigma$; standard deviation of the noise

$\beta$; stopping (regularization) parameter

$t_{max}$; maximum number of iterations.

$c\leftarrow0$; the projection coefficient

$c_{max}\leftarrow0$; the selected projection coefficient

$k$; the index of the selected column

for($t=0,1,...,t_{max}-1$)

\quad{}\{

\quad{}$c_{max}\leftarrow0$;

\quad{}for($j=0,1,...,J-1$)

\quad{}\quad{}\{

\quad{}\quad{}$c\leftarrow\left\langle r,\Gamma^{(j)}\right\rangle \left\Vert \Gamma^{(j)}\right\Vert _{2}^{-1}$;

\quad{}\quad{}if($\left|c\right|\geq\left|c_{max}\right|$)

\quad{}\quad{}\quad{}\{

\quad{}\quad{}\quad{}$c_{max}\leftarrow c$;

\quad{}\quad{}\quad{}$k\leftarrow j$;

\quad{}\quad{}\quad{}\}

\quad{}\quad{}\}

\quad{}$c\leftarrow c_{max}\left\Vert \Gamma^{(k)}\right\Vert _{2}^{-1}$;

\quad{}$\xi_{k}\leftarrow\xi_{k}+c$;

\quad{}$r\leftarrow r-c\Gamma^{(k)}$;

\quad{}if($\left\Vert r\right\Vert _{2}^{2}\leq(\beta\sigma)^{2}$)
then break;

\quad{}\}

return $\xi$;


\clearpage



\clearpage



\begin{figure}
\epsscale{1.00}
\plotone{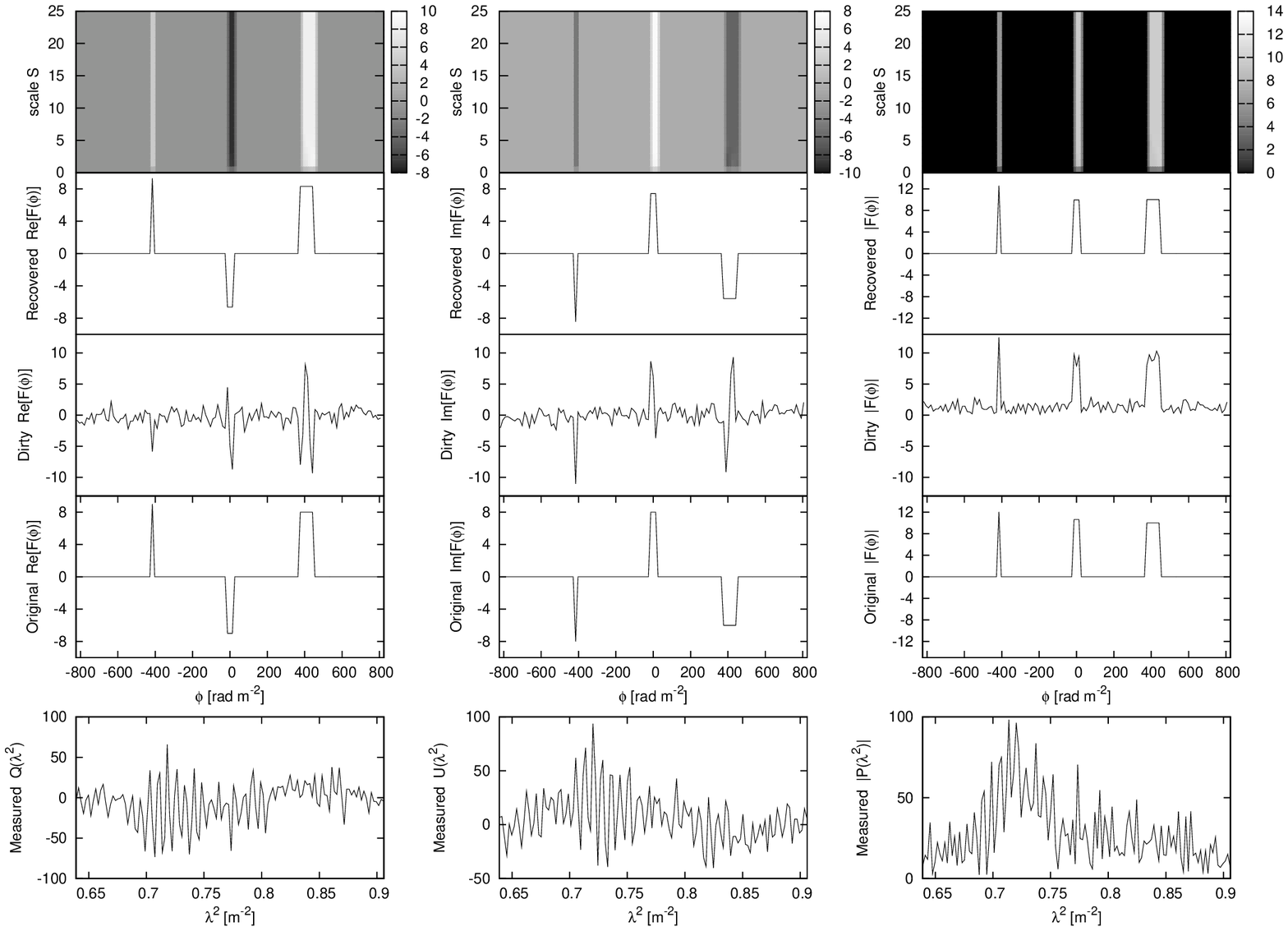}
\caption{WSRT experiment layout, noiseless exact sampling case:
$M=N=126$ and $\phi_{R}=\delta\phi=12.990\,\mathrm{rad}\,\mathrm{m}^{-2}$. 
The figure is bottom-up organized: the bottom row is the measured data, i.e. 
$Q(\lambda^2)$, $U(\lambda^2)$, and $P(\lambda^2)$; the second row is the 
input (original) model of $F(\phi)$; the third row is the dirty $F(\phi)$; the forth row is the RM-MP algorithm recovered $F(\phi)$; and  
the fifth row is the multi-scale representation of the solution (see text for details). \label{fig1}}
\end{figure}

\clearpage

\begin{figure}
\epsscale{1.00}
\plotone{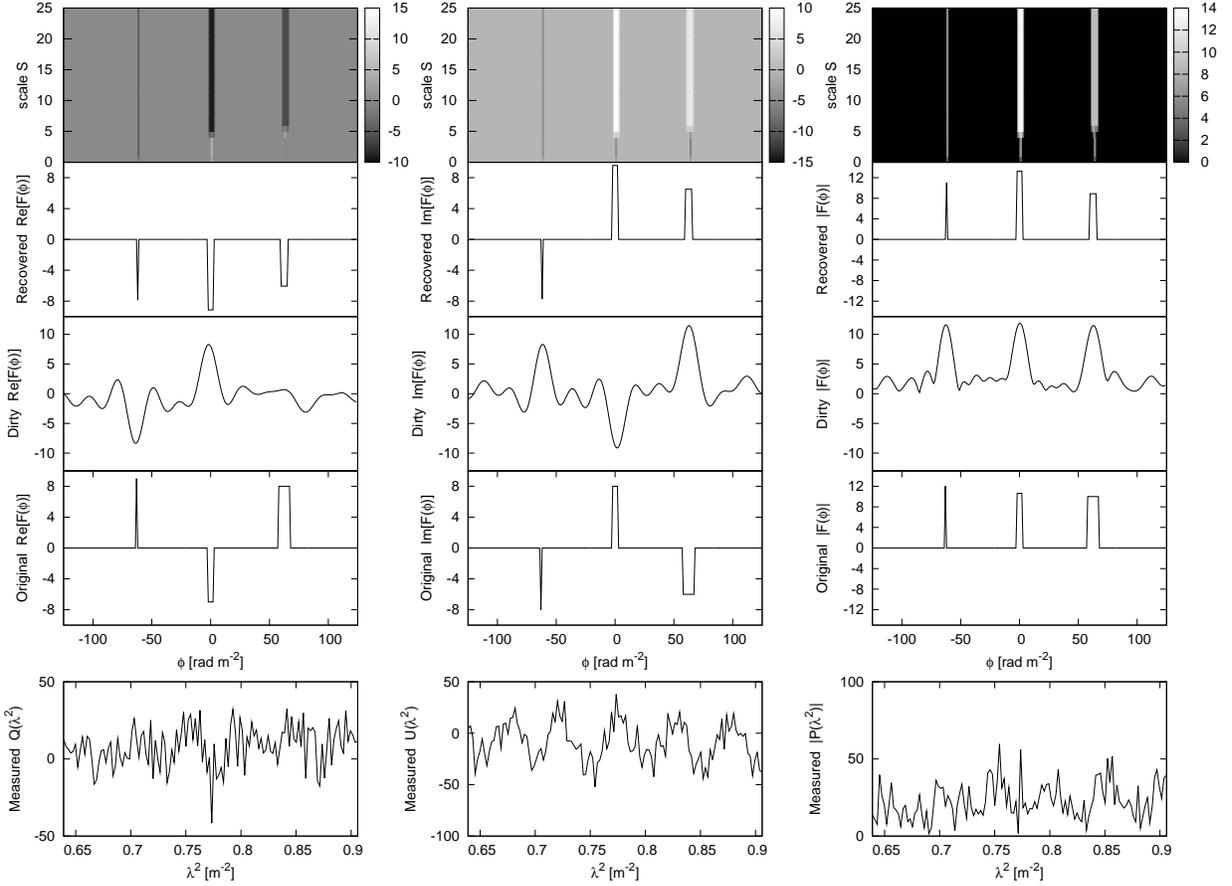}
\caption{WSRT experiment layout, noisy sampling case ($\sigma=\sqrt{N}$):
$N=126$, $M=252$ and $\phi_{R}=1\,\mathrm{rad}\,\mathrm{m}^{-2}\ll\delta\phi$.\label{fig2}}
\end{figure}

\clearpage

\begin{figure}
\epsscale{1.00}
\plotone{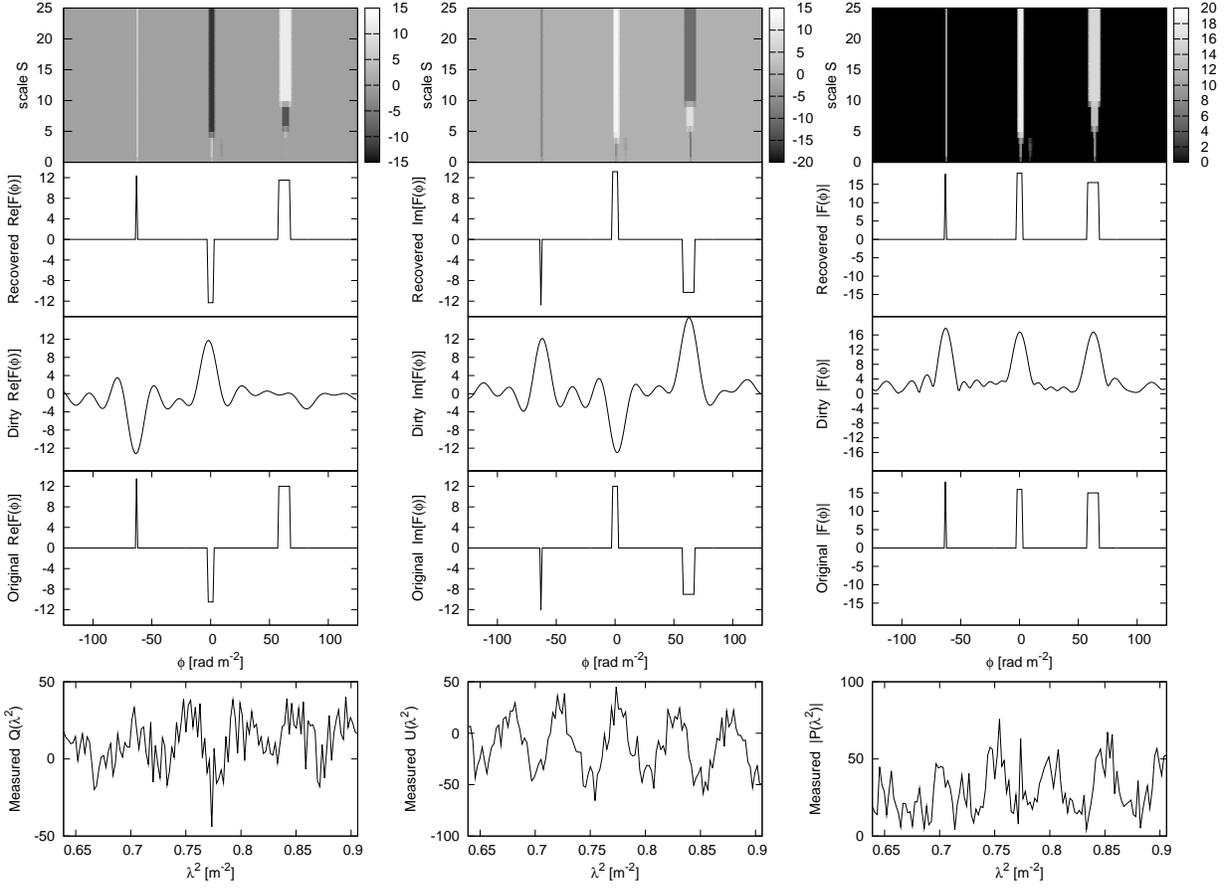}
\caption{WSRT experiment layout, noisy sampling case ($\sigma=\sqrt{N}$),
with a higher signal to noise ratio: $N=126$, $M=252$ and $\phi_{R}=1\,\mathrm{rad}\,\mathrm{m}^{-2}\ll\delta\phi$.\label{fig3}}
\end{figure}

\clearpage

\begin{figure}
\epsscale{1.00}
\plotone{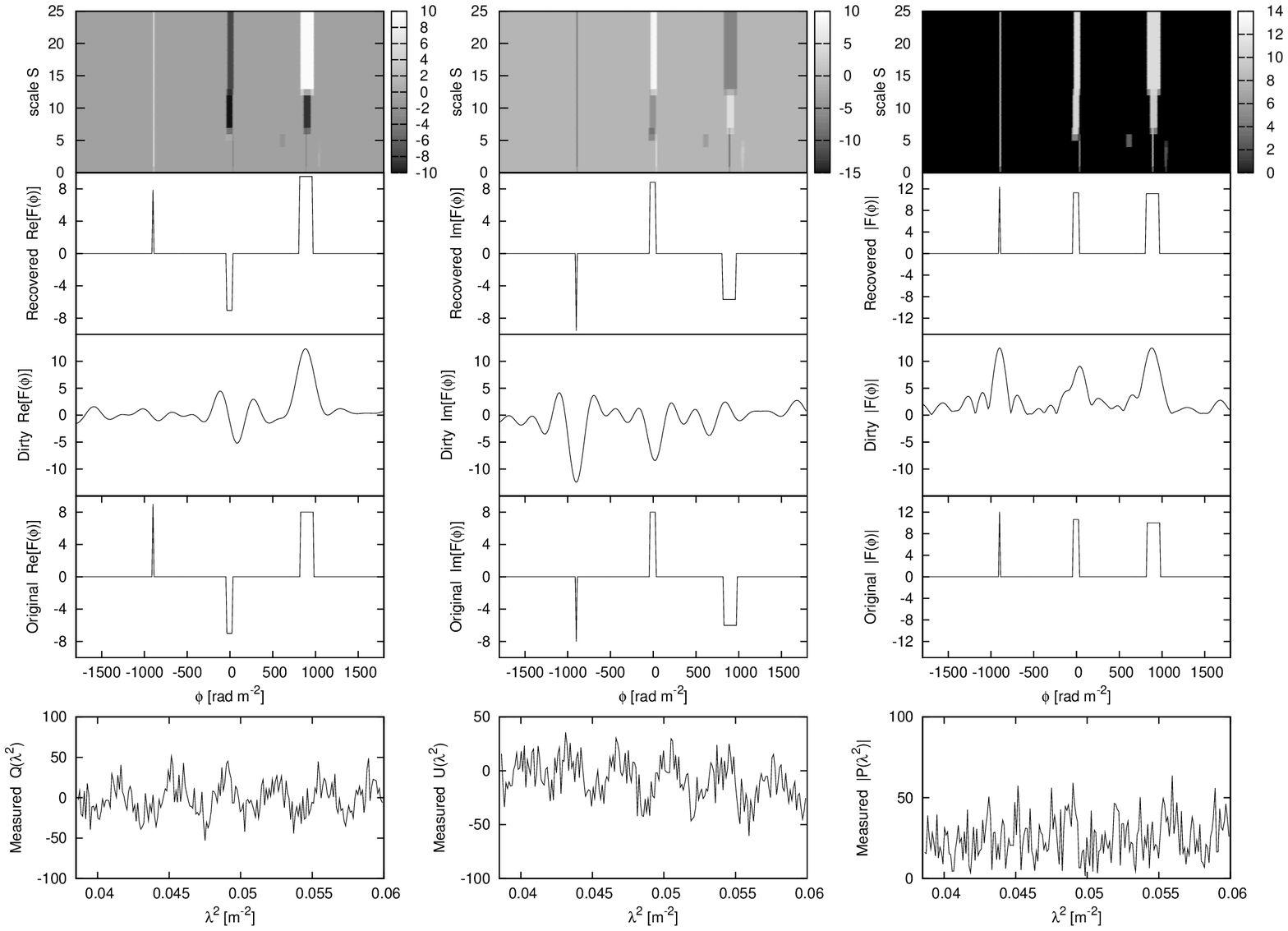}
\caption{Arecibo experiment layout, noisy sampling case ($\sigma=\sqrt{N}$):
$N=200$, $M=300$ and $\phi_{R}=12\,\mathrm{rad}\,\mathrm{m}^{-2}\ll\delta\phi$.\label{fig4}}
\end{figure}

\begin{figure}
\epsscale{1.00}
\plotone{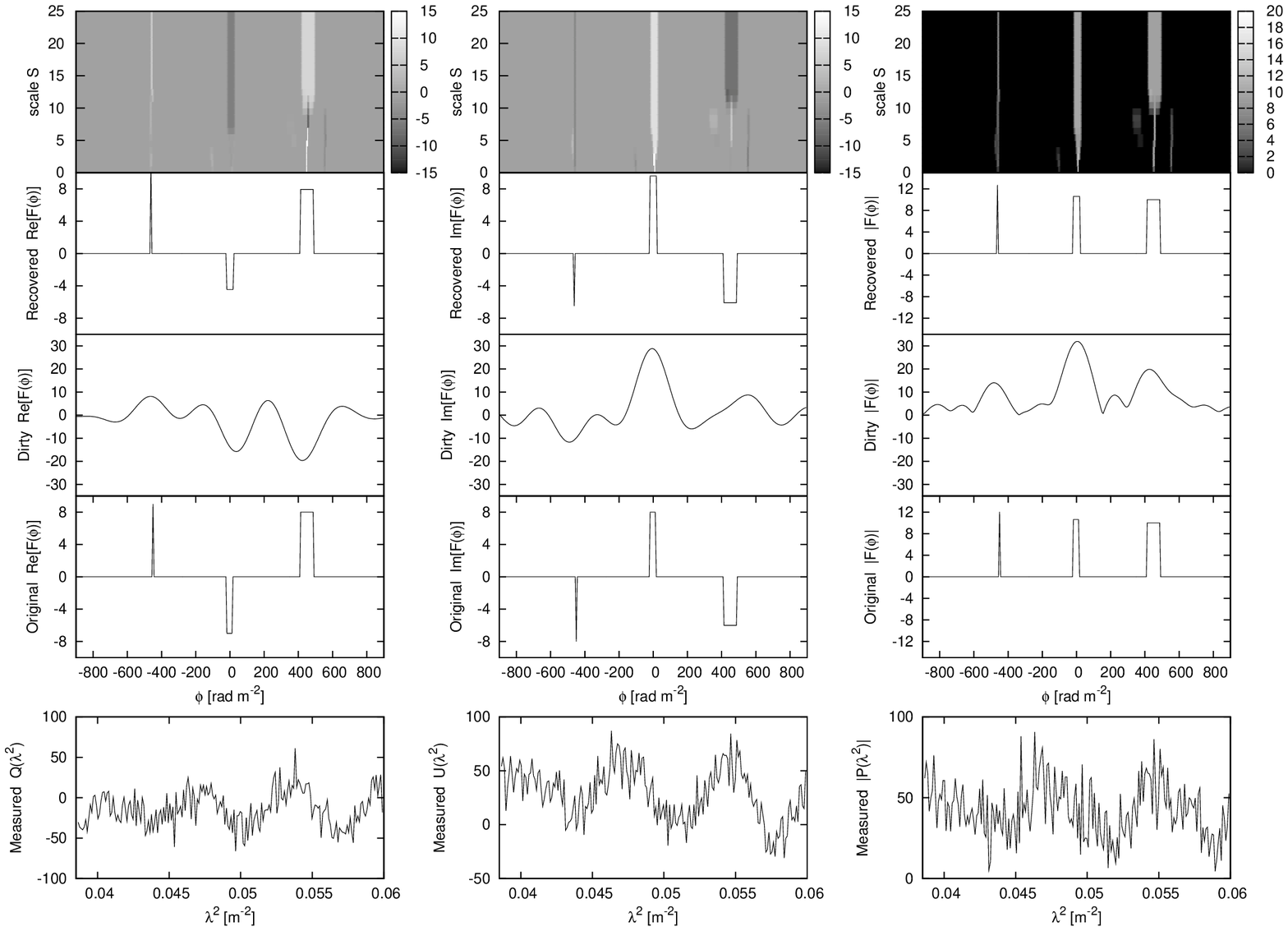}
\caption{Arecibo experiment layout, noisy sampling case ($\sigma=\sqrt{N}$):
$N=200$, $M=300$ and $\phi_{R}=6\,\mathrm{rad}\,\mathrm{m}^{-2}\ll\delta\phi$.\label{fig5}}
\end{figure}

\clearpage

\begin{figure}
\epsscale{1.00}
\plotone{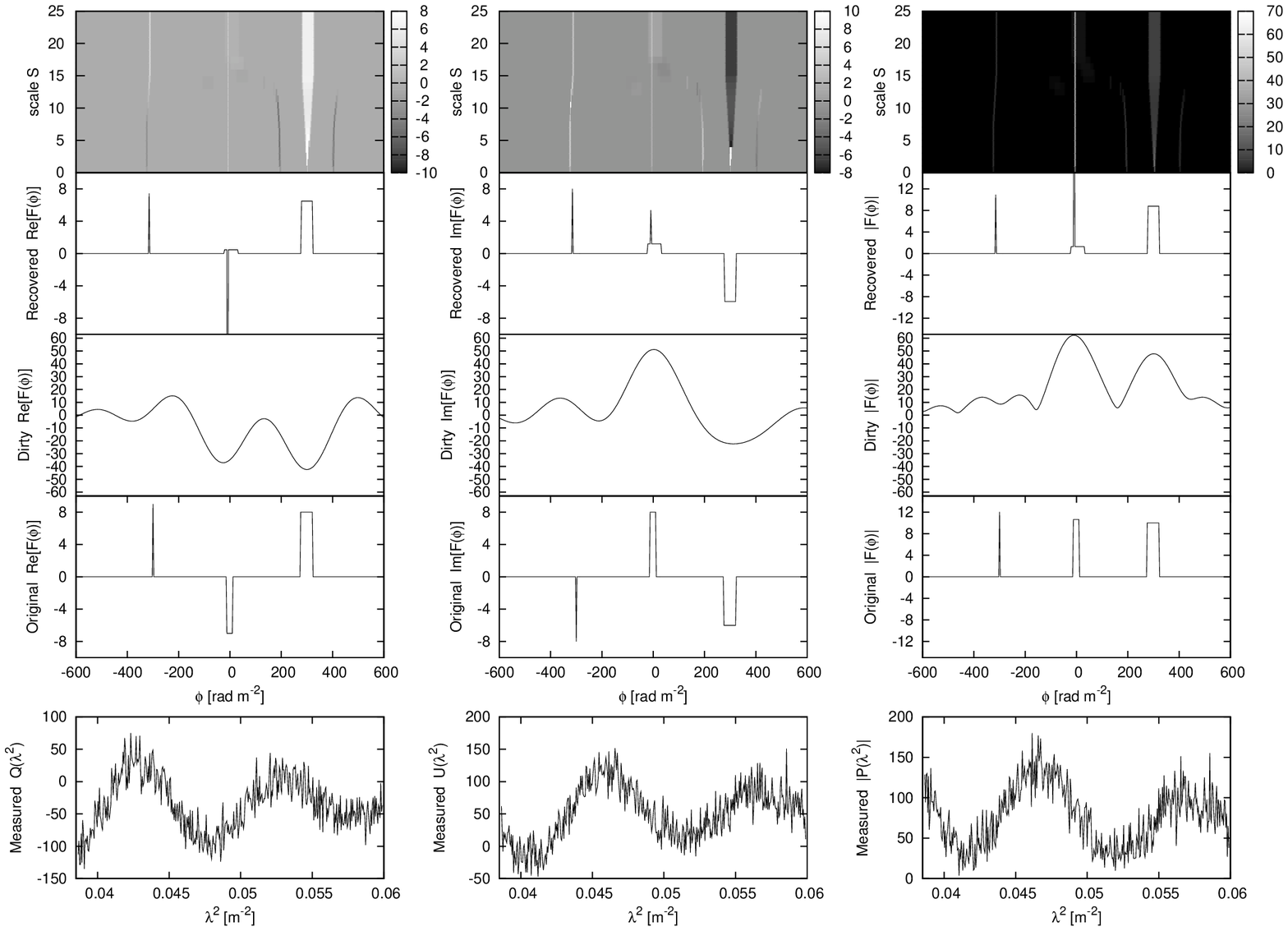}
\caption{Arecibo experiment layout, noisy sampling case ($\sigma=\sqrt{N}$):
$N=M=400$, and $\phi_{R}=3\,\mathrm{rad}\,\mathrm{m}^{-2}\ll\delta\phi $.\label{fig6}}
\end{figure}

\clearpage

\begin{figure}
\epsscale{1.00}
\plotone{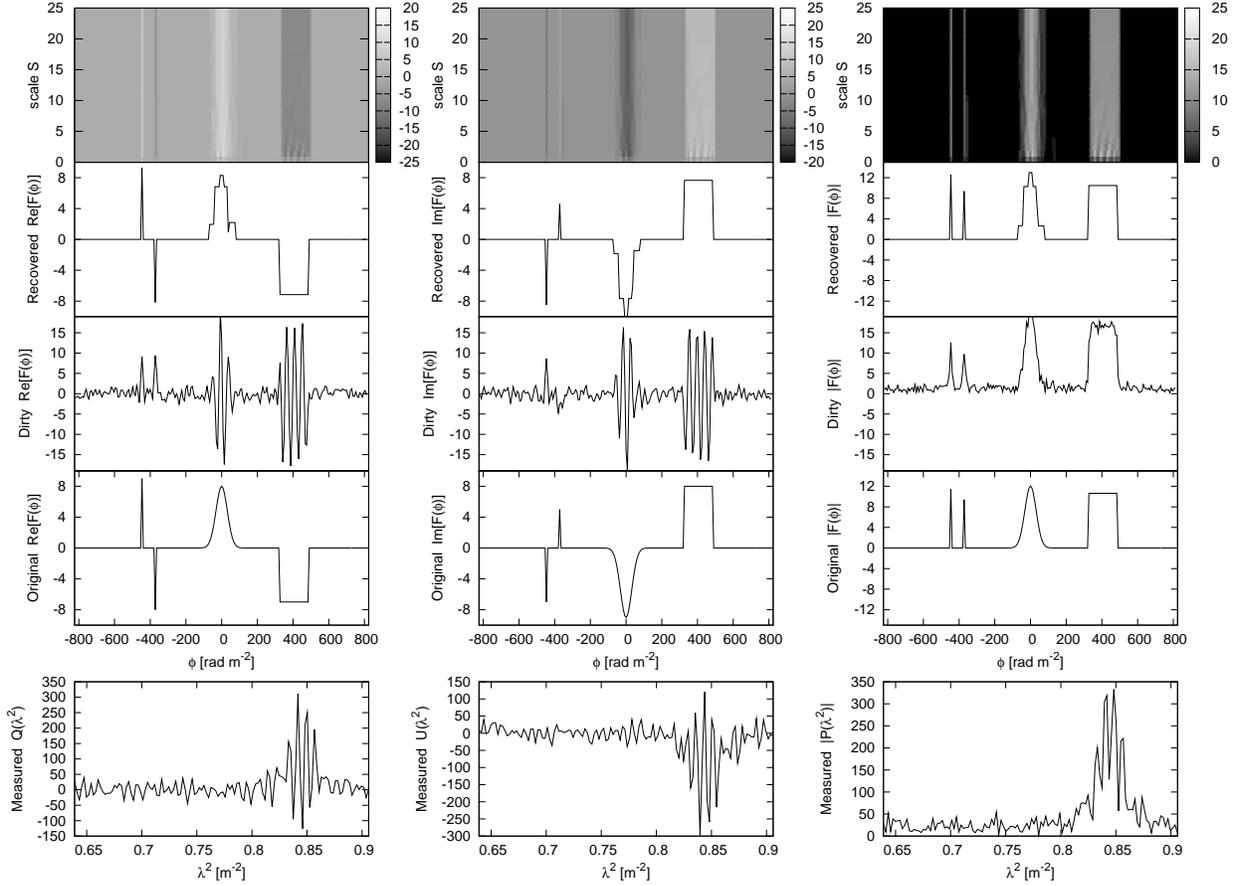}
\caption{WSRT experiment layout, noisy sampling case ($\sigma=\sqrt{N}$):
$N=126$, $M=220$ and $\phi_{R}=7.440\,\mathrm{rad}\,\mathrm{m}^{-2}=0.57\delta\phi $.\label{fig7}}
\end{figure}

\clearpage

\begin{figure}
\epsscale{0.60}
\plotone{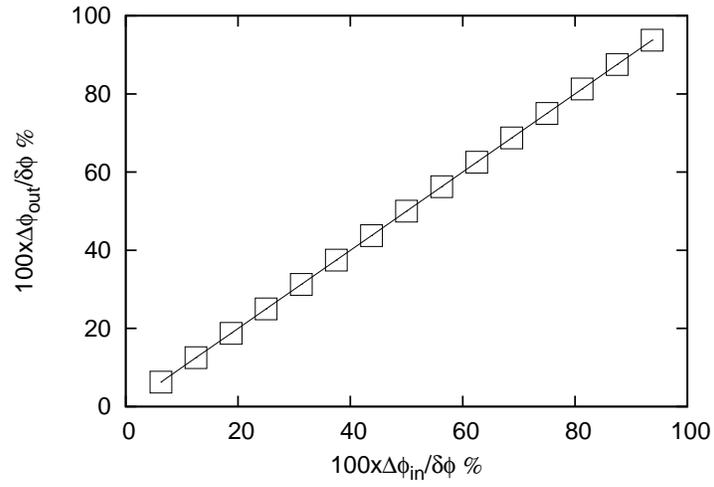}
\caption{The width of the boxcar function response $\Delta\phi_{out}$ as a function of the separation width $\Delta\phi_{in}$ between two Dirac components.\label{fig8}}
\end{figure}

\clearpage

\begin{figure}
\epsscale{1.00}
\plotone{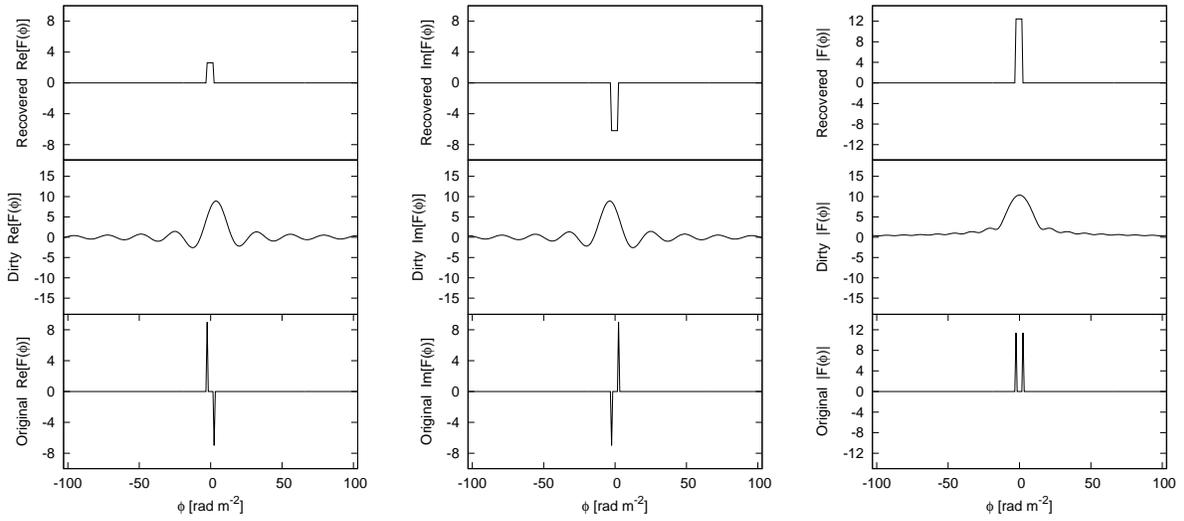}
\caption{A typical response of the RM-MP algorithm for two Dirac components separated 
by $\Delta\phi_{in}=5\phi_{R}=4.06 \mathrm{rad}\,\mathrm{m}^{-2} < \delta\phi = 12.99 \mathrm{rad}\,\mathrm{m}^{-2}$. 
WSRT experiment layout, noiseless sampling case: $N=126$, $M=1008$ and $\phi_{R}=0.812\,\mathrm{rad}\,\mathrm{m}^{-2}$.\label{fig9}}
\end{figure}

\end{document}